\documentclass[twocolumn,showpacs]{revtex4}
\usepackage{amsmath}
\usepackage{amssymb}
\usepackage{epsfig}
\usepackage{pstricks}

\def\Unitmatrix{\mbox{1\hspace{-.25em}I}}

\begin{document}

\title{Strict site-occupation constraint in 2d Heisenberg models
\\ and dynamical mass generation in $QED_3$ at finite temperature.}
\author{Raoul Dillenschneider}
\email[E-mail address : ]{rdillen@lpt1.u-strasbg.fr}
\author{Jean Richert}
\email[E-mail address : ]{richert@lpt1.u-strasbg.fr}
\affiliation{
Laboratoire de Physique Th\'eorique, UMR 7085 CNRS/ULP,\\
Universit\'e Louis Pasteur, 67084 Strasbourg Cedex,\\ 
France} 

\date{\today}

\begin{abstract}
We study the effect of site occupation in $2d$ quantum spin systems at finite 
temperature in a $\pi$-flux state description at the mean-field level. We 
impose each lattice site to be occupied by a single $SU(2)$ spin. This is 
realized by means of a specific prescription. We consider the low-energy 
Hamiltonian which is mapped into a $QED_3$ Lagrangian of spinons. We compare 
the dynamically generated mass to the one obtained by means of an average site 
occupation constraint. 

\end{abstract}


\maketitle

\section{Introduction}

Quantum Electrodynamics $QED_{(2+1)}$ is a common framework aimed to 
describe strongly correlated systems such as quantum spin systems in $1$ 
time and $2$ space dimensions, as well as related specific phenomena 
like high-$T_c$ superconductivity \cite{GhaemiSenthil-05, Morinari-05, 
LeeNagaosaWen-04}. A gauge field formulation of antiferromagnetic Heisenberg 
models in $d = 2$ dimensions leads to a $QED_3$ action for spinons,
see f. i. Ghaemi and Senthil  \cite{GhaemiSenthil-05} and Morinari 
\cite{Morinari-05}. This description raises the problem of the
mean-field solution and the correlated question of the confinement of
test charges which leads to the impossibility to determine the quantum 
fluctuation contributions through a loop expansion in this approach 
\cite{HandsKogutLucini, Herbut-02, NogueiraKleinert-05}.

We consider here the $\pi$-flux state approach introduced by Affleck and 
Marston \cite{aff1,aff2}. The occupation of sites of the system by a single
particle is generally introduced by means of a Lagrange multiplier procedure
\cite{aue,awe}. In the present work we implement a strict site-occupation. It
can be constructed by means of constraints imposed through a specific 
projection operator which introduces an imaginary chemical potential. This has
been proposed by  Popov and Fedotov \cite{Popov-88} for $SU(2)$ spins
and generalized by Kiselev et al. \cite{KFO-01} to $SU(N)$ semi-fermionic
Hamiltonians. It is our aim in the present work to confront the outcome of 
the  two approaches.



Here we concentrate on the behaviour of the spinon mass which is generated 
dynamically by an $U(1)$ gauge field. Appelquist et al. 
\cite{Appelquist1,Appelquist2} showed that at zero temperature the originally 
massless fermion can acquire a dynamically generated mass when the number $N$ 
of fermion flavors is lower than the critical value $N_c = 32/\pi^2$. Later 
Maris \cite{Maris} confirmed the existence of a critical value $N_c \simeq 3.3$
below which the dynamical mass can be generated. Since we consider only 
spin-$1/2$ systems, $N=2$ and hence $N<N_c$.

At finite temperature Dorey and Mavromatos \cite{DoreyMavromatos} and Lee
\cite{Lee-98} showed that the dynamically generated mass vanishes at a 
temperature $T$ larger than the critical one $T_c$.

We shall show below that the imaginary chemical potential introduced by Popov 
and Fedotov \cite{Popov-88} modifies noticeably the effective potential between
two charged particles and doubles the dynamical mass transition temperature, 
in agreement with former work at the same mean-field level \cite{PRB}.

The outline of the paper is the following. In section \ref{sectionPF}
we recall the projection procedure introduced by Popov and Fedotov $(PFP)$ 
leading to a rigorous constraint on the lattice site occupation.
In section \ref{sectionMeanField} we derive the Lagrangian which couples 
a spinon field to a $U(1)$ gauge field. Section \ref{sectionPhotonPropagator} 
is devoted to the comparison of the effective potential constructed with and 
without strict occupation constraint. In section \ref{sectionDynamicalMass} 
we present the calculation of the mass term using the Schwinger-Dyson 
equation of the spinon.

\section{Site occupation constraint for quantum spin systems at 
finite temperature. \label{sectionPF}}

Heisenberg quantum spin Hamiltonians of the type

\begin{eqnarray}
H = \frac {1}{2}\sum_{i,j} J_{ij} \vec S_{i} \vec S_{j}
\label{eq1} 
\end{eqnarray}
with $\{J_{ij}\} > 0$ can be projected onto Fock space by means 
of the transformation  

\begin{eqnarray}
S^{+}_{i} &=& f^{\dagger}_{i, \uparrow} f_{i, \downarrow}
\notag \\
S^{-}_{i} &=& f^{\dagger}_{i, \downarrow} f_{i, \uparrow}
\notag \\
S^{z}_{i} &=& \frac{1}{2} (f^{\dagger}_{i,\uparrow} f_{i,\uparrow} - 
f^{\dagger}_{i,\downarrow} f_{i,\downarrow})
\label{eq2}
\end{eqnarray}
where $\{f_{i, \sigma}, f_{i, \sigma}\}$ are anticommuting 
fermion operators which create and annihilate spinon with $\sigma=\pm 1/2$.

This transformation is not bijective because the dimensionality of 
Fock space is larger than the dimensionality of the space in which 
the spin operators $\{\vec S_{i}\}$ are acting. Indeed, in Fock space, 
each site $i$ can be occupied by $0$, $1$ or $2$ fermions corresponding 
to the states $|0,0>,|1,0>,|0,1>,|1,1>$ where $|0,0>$ is the particle 
vacuum, $|1,0> = |+ 1/2>$, $|0,1> = |- 1/2>$ and $|1,1> = |+ 1/2, - 1/2>$ 
in terms of spin $1/2$ projections. Since one wants to keep states with 
one fermion per site the states $|0,0>$ and $|1,1>$ have to be 
eliminated. This can be performed on the partition function for a system at 
inverse temperature $\beta$

\begin{eqnarray*}
\mathcal{Z} = Tr \left[ e^{-\beta H} \right]
\end{eqnarray*}
where the trace is taken over the whole Fock space by the introduction of a 
projection operator

\begin{eqnarray*}
\mathcal{Z} = Tr \left[ e^{-\beta (H - \mu N)} \right] 
\end{eqnarray*}
where $N$ is the particle number operator and $\mu = i \pi/ 2 \beta$ 
an imaginary chemical potential \cite{Popov-88}. Indeed, the presence of the 
states $|0,0>_{i}$ and $|1,1>_{i}$ on site $i$ leads in  $Z$ to  
phase contributions which eliminate each other 

\begin{eqnarray*}
e^{i*0} + e^{i* \pi} = 0 
\label{eq3} 
\end{eqnarray*}
and hence the contributions of these spurious states are cancelled as a whole.

The common alternative approximate projection procedure would be to introduce 
a chemical potential in terms of $real$ Lagrange multipliers 
$\{\lambda_{i}\}$

\begin{eqnarray*}
\mathcal{Z} = Tr \left[ e^{-\beta H} {\prod_{i} \int d\lambda_{i}  
e^{\lambda_{i}(n_{i}-1)}} \right]
\label{eq4}
\end{eqnarray*}
where $n_{i}$ is the particle number operator on site $i$ and the
$\{\lambda_{i}\}$ are fixed by means of a saddle point procedure.

\section{Spin state mean-field ansatz in 2d \label{sectionMeanField}}

In 2$d$ space the Heisenberg Hamiltonian given by Eq.\eqref{eq1} can 
be written in terms of composite non-local operators $\{{\cal D}_{ij}\}$ 
("diffusons") \cite{aue} defined as 

\begin{eqnarray*}
{\mathcal D}_{ij} = f_{i, \uparrow}^{\dagger} f_{j, \uparrow} + 
f_{i, \downarrow}^{\dagger} f_{j, \downarrow}
\end{eqnarray*}

If the coupling strengths are fixed as 

\begin{eqnarray*}
J_{ij} = J 
\underset{\vec{\eta}}{\sum} \delta \left(\vec{r}_i - \vec{r}_j \pm \vec{\eta} 
\right)
\end{eqnarray*}
where $\vec{\eta}$ is a lattice vector $\{a_1,a_2\}$ in the $\vec {Ox}$  
and $\vec {Oy} $ directions the Hamiltonian takes the form

\begin{eqnarray}
H = - J \sum_{<ij>} (\frac{1}{2} {\cal D}^{\dagger}_{ij} {\cal D}_{ij} - 
\frac{n_i}{2} + \frac{n_i  n_j}{4})  
\label{eq21} 
\end{eqnarray}
where $i$ and $j$  are nearest neighbour sites.

The number operator products $\{n_i  n_j\}$ in Eq.\eqref{eq21} are quartic 
in terms of creation and annihilation operators in Fock space. 
In principle the formal treatment of these terms requires the introduction
of a mean field procedure.
One can however show that the presence of this term has no
influence on the results obtained from the partition function. As a 
consequence we leave it out from the beginning as well as the contribution 
corresponding to the $\{n_i\}$ terms.

\subsection{Exact occupation procedure}

Starting with the Hamiltonian

\begin{eqnarray}
H = - \frac{J}{2} \sum_{<ij>}   {\mathcal D}^{\dagger}_{ij} {\cal D}_{ij} 
- \mu N 
\label{eq22} 
\end{eqnarray}
the partition function $Z$ can be written in the form

\begin{eqnarray*}
\mathcal{Z} = \int \prod_{i, \sigma}{\mathcal{D}} 
(\{\xi^{*}_{i, \sigma},\xi_{i,\sigma}\})
e^{- \, A(\{\xi^{*}_{i,\sigma},\xi_{i,\sigma}\})} 
\label{eq5}
\end{eqnarray*}
where the $\{\xi^{*}_{i,\sigma},\xi_{i,\sigma}\}$ are Grassmann variables 
corresponding to the operators $\{f^{\dagger}_{i \sigma}, f_{i \sigma}\}$ 
defined above. They depend on the imaginary time $\tau$ in the interval 
$[0,\beta]$. In the continuum limit the action $A$ is given by

\begin{eqnarray*}
A(\{\xi^{*}_{i,\sigma},\xi_{i,\sigma}\}) &=& \int_{0}^{\beta} d\tau \Bigg(
\sum_{i,\sigma} \xi^{*}_{i,\sigma} (\tau) \partial_{\tau} \xi_{i,\sigma} 
(\tau) 
\notag \\
&+& \mathcal{H}(\{\xi^{*}_{i,\sigma} (\tau),\xi_{i,\sigma} 
(\tau)\}) \Bigg)
\label{eq6}
\end{eqnarray*}
where 

\begin{eqnarray}
\mathcal{H} (\tau) = H(\tau) - \mu N(\tau)
\label{eq7}
\end{eqnarray}
and $N(\tau)$ is the total particle number operator. 
A Hubbard-Stratonovich transformation on the corresponding functional integral 
partition function in which the action contains the occupation number 
operator as seen in Eq.\eqref{eq7} eliminates the quartic contributions
generated by Eq.\eqref{eq2} and introduces the mean fields $\{\Delta_{ij}\}$.
The Hamiltonian takes then the form

\begin{eqnarray}
\mathcal{H} = \frac{2}{|J|}\underset{<ij>}{\sum} \bar{\Delta}_{ij} \Delta_{ij} 
+\underset{<ij>}{\sum} \left[ \bar{\Delta}_{ij} {\cal D}_{ij} +
\Delta_{ij} {\cal D}^{\dagger}_{ij} \right]  - \mu N
\notag \\
\label{eq23}
\end{eqnarray}

The fields $\{\Delta_{ij}\}$ and their complex conjugates $\bar \Delta_{ij}$ 
can be decomposed into a mean-field contribution and a fluctuation term
 
\begin{eqnarray*}
\Delta_{ij} = \Delta_{ij}^{mf} + \delta \Delta_{ij}
\label{eq24}
\end{eqnarray*}

The field $\Delta_{ij}^{mf}$ can be chosen as a complex quantity 
$\Delta_{ij}^{mf}=|\Delta_{ij}^{mf}|e^{i\phi_{ij}^{mf}}$. 

The phase $\phi_{ij}^{mf}$ is fixed in the following way. Consider a 
square plaquette 
$\Box \equiv (\vec i, \vec i + \vec {e}_{x},\vec i + \vec {e}_{y},
\vec i + \vec {e}_{x} + \vec {e}_{y})$ where $\vec {e}_{x}$ and $\vec {e}_{y}$ 
are the unit vectors along the directions $\vec {Ox}$ and $\vec {Oy}$ 
starting from site $\vec i$ on the lattice. On this plaquette we define 

\begin{eqnarray*}
\phi = \prod_{(ij) \in \Box}\phi_{ij}^{mf}
\end{eqnarray*}
which is taken to be constant. If the gauge phase $\phi_{ij}^{mf}$ fluctuates 
in such a way that $\phi$ stays constant the average of $\Delta_{ij}^{mf}$ 
will be equal to zero in agreement with Elitzur's theorem \cite{el1}. 
In order to guarantee the $SU(2)$ invariance of the mean-field Hamiltonian
along the plaquette we follow \cite{wen,aff1,aff2,awe,lee} 
and introduce

\begin{eqnarray*}
\phi_{ij}=
\begin{cases}
e^{i.\frac{\pi}{4}(-1)^i}, \text{if } \vec{r}_j=\vec{r}_i+\vec{e}_x \\
e^{-i.\frac{\pi}{4}(-1)^i}, \text{if } \vec{r}_j=\vec{r}_i+\vec{e}_y \\
\end{cases}
\end{eqnarray*}
where $\vec{e}_x$ and $\vec{e}_y$ join the site $i$ to its nearest neighbours
$j$. Then the total flux through the fundamental plaquette is such that 
$\phi = \pi$ which guarantees that the $SU(2)$ symmetry of the plaquette is
respected \cite{Marston} .

At the mean-field level the partition function reads 

\begin{eqnarray*}
\mathcal{Z}_{mf} = e^{- \beta (\mathcal{H}_{mf} - \mu N)}
\label{eq27}
\end{eqnarray*}
where 

\begin{eqnarray}
\mathcal{H}_{mf} &=& \frac{2}{|J|}\underset{<ij>}{\sum} 
\bar \Delta_{ij}^{mf}.\Delta_{ij}^{mf} 
\notag \\
&+& \underset{<ij>}{\sum} \left[
\bar \Delta_{ij}^{mf} {\cal D}_{ij} + 
\Delta_{ij}^{mf} {\cal D}^{\dagger}_{ij} \right]  - \mu N
\label{eq28}
\end{eqnarray}
as read immediatly from Eq. \eqref{eq23}

\begin{figure}
\center
\begin{pspicture}(4,4)
\psframe(0,0)(3,3)
\psline{->}(-0.5,1.5)(4,1.5)
\rput(4,1.2){\text{$k_x$}}
\psline{->}(1.5,-0.5)(1.5,4)
\rput(1.2,4){\text{$k_y$}}
\pspolygon*(0,1.5)(1.5,3)(3,1.5)(1.5,0)
\rput(3.3,1.7){\text{$+\pi$}}
\rput(1.8,3.3){\text{$+\pi$}}
\rput(1.8,-0.2){\text{$-\pi$}}
\rput(-0.4,1.7){\text{$-\pi$}}
\end{pspicture}
\caption{The two-dimensional spin Brillouin Zone (black area) and the
lattice Brillouin Zone (whole square).}
\label{Fig1}
\end{figure}
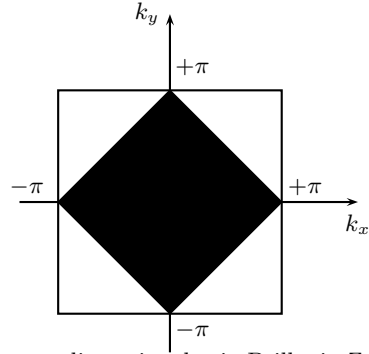

\noindent
After a Fourier transformation the Hamiltonian \eqref{eq28} takes the form

\begin{eqnarray}
\mathcal{H}_{mf} &=& \mathcal{N} z \frac{\Delta^2}{|J|}
\notag \\
&+& \underset{\vec{k} \in SBZ}{\sum} \underset{\sigma}{\sum}
\left(
f^\dagger_{\vec{k},\sigma} \,
f^\dagger_{\vec{k}+\vec{\pi},\sigma}
\right)
\left[
\widetilde{H}
\right]
\left(
\begin{array}{c}
f_{\vec{k},\sigma} \\
f_{\vec{k}+\vec{\pi},\sigma}
\end{array}
\right)
\label{HDijK}
\end{eqnarray}
with

\begin{eqnarray*}
\left[ \widetilde{H} \right] =
\left[
\begin{array}{cc}
-\mu + \Delta \cos \frac{\pi}{4} z \gamma_{k_x,k_y} &
-i \Delta \sin \frac{\pi}{4} z \gamma_{k_x,k_y+\pi} \\
+i \Delta \sin \frac{\pi}{4} z \gamma_{k_x,k_y+\pi} &
-\mu - \Delta \cos \frac{\pi}{4} z \gamma_{k_x,k_y} 
\end{array}
\right]
\end{eqnarray*}
where $\Delta \equiv |\Delta^{mf}|$. The Spin Brillouin Zone (SBZ) covers
half of the Brillouin Zone (see figure \ref{Fig1}). The $\gamma_{\vec{k}}$'s 
are defined by

\begin{eqnarray*}
\gamma_{\vec{k}} = \frac{1}{z} \underset{\vec{\eta}}{\sum} 
e^{i \vec{k}.\vec{\eta}} = \frac{1}{2} \left( \cos k_x a_1  + \cos k_y a_2 
\right)
\end{eqnarray*}

\subsection{The $\pi$-flux Dirac action}

As already shown in earlier work by Ghaemi and Senthil 
\cite{GhaemiSenthil-05} and Morinari \cite{Morinari-05} the spin liquid
Hamiltonian \eqref{HDijK} for systems at low energy can be described by 
four-component Dirac spinons in the continuum limit. 
The Dirac action of spin liquid in (2+1) dimensions is derived in appendix
\ref{AppendixA}.
In Euclidean space this action reads

\begin{eqnarray}
S_{E} = \int_0^\beta d\tau \int d^2\vec{r} \underset{\sigma}{\sum}
\bar{\psi}_{\vec{r} \sigma} \left[ \gamma^0 \left( \partial_\tau - \mu \right)
+ \widetilde{\Delta} \gamma^k \partial_k \right] \psi_{\vec{r} \sigma}
\notag \\
\label{SpinonAction}
\end{eqnarray}
where $\widetilde{\Delta} = 2 \Delta \cos \frac{\pi}{4}$ is the ``light 
velocity'', and $\{\gamma^{\mu}\}$ are the Dirac gamma matrices in (2+1) 
dimensions. Spinons move in a ``gravitational'' field and the metric can be 
handled in a Minkowskian (or Euclidean) metric \cite{Volovik} assuming 
$\widetilde{\Delta} = 1$ without altering the physics of the problem.

Since the Heisenberg Hamiltonian \eqref{eq1} is gauge invariant 
in the transformation $\psi \rightarrow e^{i g a_\mu} \psi$ the Dirac 
action has to be written in the form

\begin{eqnarray}
S_{E} = \int_0^\beta \int d^2\vec{r}
&\Bigg\{&
- \frac{1}{2} a_\mu \left[ \left(\Box \delta^{\mu \nu}
+ (1 - \lambda) \partial^\mu \partial^\nu \right) \right] a_\nu
\notag \\
&+& \underset{\sigma}{\sum}
\bar{\psi}_{\vec{r} \sigma} \left[
\gamma_\mu \left( \partial_\mu - i g a_\mu \right)
\right] \psi_{\vec{r} \sigma}
\Bigg\}
\label{SpinonGaugeAction}
\end{eqnarray}
Here $g$ is the coupling strength
between the gauge field $a_\mu$ and the Dirac spinons
$\psi$. In \eqref{SpinonGaugeAction} the first term corresponds to the 
``Maxwell'' term $-\frac{1}{4}f_{\mu \nu} f^{\mu \nu}$ of the gauge field 
$a_\mu$ where $f^{\mu \nu} = \partial_\mu a_\nu - \partial_\nu a_\mu$,
$\lambda$ is the parameter of the Faddeev-Popov gauge fixing term
$-\lambda \left(\partial^\mu a_\mu \right)^2$ \cite{Itzykson} and  
$\delta^{\mu \nu}$ the Kronecker $\delta$.
$\Box = \partial_\tau^2 + \vec{\nabla}^2$ is the Laplacian in Euclidean 
space-time. This form of the action originates from a shift of the imaginary 
time derivation $\partial_\tau \rightarrow \partial_\tau + \mu$ and leads to 
a new definition of the Matsubara frequencies only for the fermion fields 
$\psi$ \cite{Popov-88} which read then

\begin{eqnarray*}
\widetilde{\omega}_{F,n} = \omega_{F,n} - \mu/i = \frac{2 \pi}{\beta} (n + 1/4)
\end{eqnarray*}
  
This modification will induce substantial consequences as it will be shown 
in the following.

\section{The ``Photon'' propagator at finite temperature
\label{sectionPhotonPropagator}}

Integrating over the fermion fields $\psi$ leads to a pure gauge Lagrangian
$\mathcal{L}_a = \frac{1}{2} a_\mu \Delta_{\mu \nu}^{-1} a_\nu$ where
$\Delta_{\mu \nu}$ is the dressed photon propagator from which we shall  
extract an effective interaction potential $V(R)$ between two test particles 
and extract a dynamically generated fermion mass. 

The finite-temperature photon propagator in Euclidean space (imaginary time 
formulation) verifies the Dyson equation

\begin{eqnarray}
\Delta_{\mu \nu}^{-1} &=& {\Delta_{\mu \nu}^{(0)}}^{-1} + \Pi_{\mu \nu}
\label{Dyson}
\end{eqnarray}
The detailed calculation of the polarisation function $\Pi_{\mu \nu}$
is given in appendix \ref{AppendixPolarisation}.

Since the system is at finite temperature and ``relativistic'' covariance
should be kept the polarisation function may be put in the form  \cite{Das}

\begin{eqnarray*}
\Pi_{\mu \nu} = \Pi_A A_{\mu \nu} + \Pi_B B_{\mu \nu}
\end{eqnarray*}
where $\Pi_A$ and $\Pi_B$ are related to $\widetilde{\Pi}_k$ by 
$\Pi_A = \widetilde{\Pi}_1 + \widetilde{\Pi}_2$ and 
$\Pi_B = \widetilde{\Pi}_3$. 
The expressions of $\widetilde{\Pi}_1$, $\widetilde{\Pi}_2$ and
$\widetilde{\Pi}_3$ are explicitly worked out in appendix 
\ref{AppendixPolarisation}. $A_{\mu \nu}$ and $B_{\mu \nu}$ generate an 
orthogonal tensor basis transversal to the photon momentum $q^\mu$ 

\begin{eqnarray*}
A_{\mu \nu} &=& \widetilde{\eta}_{\mu \nu} - \frac{\widetilde{q}_\mu 
\widetilde{q}_\nu} {\widetilde{q}^2} \\
B_{\mu \nu} &=& \frac{q^2}{\widetilde{q}^2} \bar{u}_\mu \bar{u}_\nu \\
A_{\mu \nu} + B_{\mu \nu} &=& \delta_{\mu \nu} - \frac{q_\mu q\nu}{q^2} \notag
\end{eqnarray*}
with $\widetilde{\eta}_{\mu \nu} = \delta_{\mu \nu} - \frac{q_\mu q_\nu}{q^2}$
, $\bar{u}_\mu = u_\mu - \frac{(q.u)}{q^2} q_\mu$ and 
$\widetilde{q}_\mu = q_\mu - (q.u)u_\mu$. Here $u_\mu = (1,0,0)$ is the 
three-vector of the thermal bath.

\begin{figure}
\epsfig{file=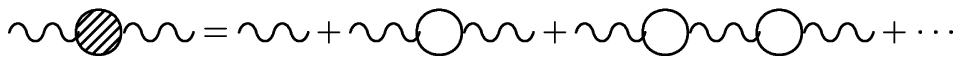,scale=0.8}
\caption{The dressed photon propagator. Wavy lines correspond to the photon
and solid loops to the fermion insertions}
\label{FeynmannGraphs1}
\end{figure}

The dressed photon propagator $\Delta_{\mu \nu}$
is obtained by the summation of the geometric series shown in figure 
\ref{FeynmannGraphs1} and reads

\begin{eqnarray}
\Delta_{\mu \nu} = \frac{A_{\mu \nu}}{q^2 + \widetilde{\Pi}_1 
+ \widetilde{\Pi}_2}
+ \frac{B_{\mu \nu}}{q^2 + \widetilde{\Pi}_3} 
- (1-1/\lambda) \frac{q_\mu q_\nu}{(q^2)^2}
\notag \\
\label{PhotonPropagator}
\end{eqnarray}

\subsection{Effective potential between test particles}

The effective static potential between two test particles of opposite chages
$g$ at distance $R$ is given by

\begin{eqnarray*}
V(R) &=& - g^2 \int_0^\beta d\tau \Delta_{00}(\tau,R)
\notag \\
&=& -g^2 \frac{1}{2\pi} \int \frac{d^2\vec{q}}{(2\pi)^2} \Delta_{00}
(q^0=0,\vec{q}) e^{i\vec{q}.\vec{R}}
\notag \\
&=& -\frac{g^2}{2\pi} \int_0^\infty dq q J_0(qR).\frac{1}{q^2 +
\widetilde{\Pi}_3(m=0)}
\end{eqnarray*}
where $J_0(qR)$ is the zero order Bessel function. The polarisation 
contribution $\widetilde{\Pi}_3(q^0=0,\vec{q})$ is equal to 
$\frac{\alpha}{\pi \beta}
\int_0^1 dx \log 2 \left( \cosh \beta q \sqrt{x(1-x)} \right)$ when taking
the $PFP$ imaginary chemical potential into account. This has to be compared
to the expression
$\frac{2 \alpha}{\pi \beta} \int_0^1 dx \log 2 \left( \cosh \frac{\beta}{2} q 
\sqrt{x(1-x)} \right)$ when the Lagrange multiplier method for
which $\lambda=0$ is used \cite{DoreyMavromatos}.

\begin{figure}
\epsfig{file=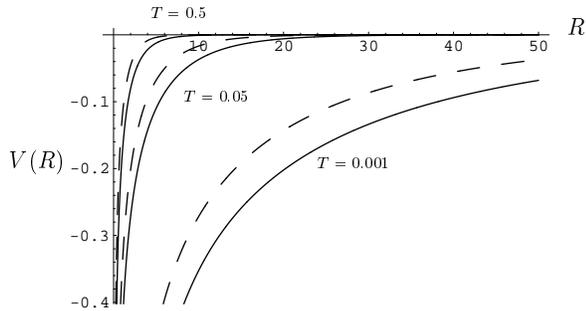,width=8cm}
\caption{Effective static potential with (full line) and without (dashed line)
the Popov-Fedotov imaginary chemical potential for the temperature 
$T=\{0.001,0.05,0.5\}$.}
\label{StaticPotential}
\end{figure}

For small momentum $q \rightarrow 0$, $\widetilde{\Pi}_3(m=0)$ can be 
identified as a mass term $(M_0^{PF}(\beta))^2$. 
For $R \gg (M_0^{PF})^{-1}$ the effective potential reads

\begin{eqnarray*}
V(R,\beta) &\simeq& - \frac{g^2}{2\pi} \int_0^\infty dq \frac{q J_0(qR)}{q^2 +
{M_0^{PF}}^2} 
\notag \\
&=& - \frac{\alpha}{N} \sqrt{\frac{1}{8 \pi R M_0^{PF}}} 
e^{-M_0^{PF} R}
\end{eqnarray*}
where $N=2$ since we consider only $S=1/2$ spins.

Figure \ref{StaticPotential} shows the effective potential between two
opposite test charges at distance $R \gg (M_0^{PF})^{-1}$. 
The screening effect is smaller when the imaginary chemical potential $\mu$
is implemented rather than the Lagrange multiplier $\lambda$.
By inspection one sees that $(M_0^{PF})^{-1} = \sqrt{2} 
(M_0^{\lambda = 0})^{-1}$.

\section{Dynamical mass generation \label{sectionDynamicalMass}}

\begin{figure}
\epsfig{file=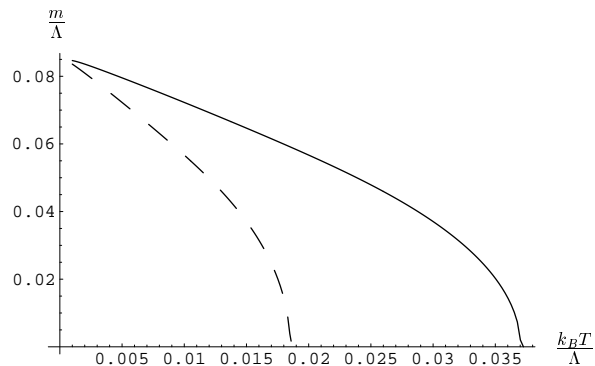,width=8cm}
\caption{Temperature dependence of the dynamical mass generated
with (full line) and without (dashed line) the use of the Popov-Fedotov 
procedure.}
\label{Fig3}
\end{figure}

We show now how the $PFP$ doubles the ``chiral'' restoring transition 
temperature of the dynamical mass generation. The Schwinger-Dyson equation 
for the spinon propagator at finite temperature reads

\begin{eqnarray}
G^{-1}(k) &=& {G^{(0)}}^{-1}(k) 
\notag \\
&-& \frac{g}{\beta} \underset{\widetilde{\omega}_{F,n}}
{\sum} \int \frac{d^2 \vec{P}}{(2 \pi)^2} \gamma_\mu G(p) \Delta_{\mu \nu}
(k-p) \Gamma_\nu
\notag \\
\label{SchwingerDyson}
\end{eqnarray}
where $p=(p_0=\widetilde{\omega}_{F,n},\vec{P})$,
$G$ is the spinon propagator, $\Gamma_\nu$ the spinon-''photon''
vertex which will be approximated here by its bare value $g \gamma_\nu$ and
$\Delta_{\mu \nu}$ is the dressed photon propagator \eqref{PhotonPropagator}.
The second term in \eqref{SchwingerDyson} is the fermion self-energy $\Sigma$,
($G^{-1} = {G^(0)}^{-1} - \Sigma$).
Performing the trace over the $\gamma$ matrices in equation 
\eqref{SchwingerDyson} leads to a self-consistent equation for the self-energy

\begin{eqnarray}
\Sigma(k) = \frac{g^2}{\beta} \underset{\widetilde{\omega}_{F,n}}{\sum} 
\int \frac{d^2 \vec{P}}{(2 \pi)^2}
\Delta_{\mu \mu}(k-p) \frac{\Sigma(p)}{p^2 + \Sigma(p)^2}
\label{selfconsitent}
\end{eqnarray}

In the low energy and momentum limit $m(\beta) = \Sigma(k) \simeq \Sigma(0)$ 
the equation \eqref{selfconsitent} simplifies to

\begin{eqnarray}
1 = \frac{g^2}{\beta} \underset{\widetilde{\omega}_{F,n}}{\sum} 
\int \frac{d^2 \vec{P}}{(2 \pi)^2}
\Delta_{\mu \mu}(-p). \frac{1}{p^2 + m(\beta)^2}
\label{Mass1}
\end{eqnarray}

\noindent
If the main contribution comes from the longitudinal part 
$\Delta_{00}(0,-\vec{P})$ of the photon propagator \eqref{Mass1} goes 
over to

\begin{eqnarray}
1 &=& \frac{g^2}{\beta} \underset{\widetilde{\omega}_{F,n}}{\sum} 
\int \frac{d^2 \vec{P}}{(2 \pi)^2}
\notag \\
&\Bigg(& \frac{1}{\vec{P}^2 + \widetilde{\Pi}_3(m=0)}. \frac{1}
{\widetilde{\omega}_{F,n}^2 + \vec{P}^2 + m(\beta)^2} \Bigg)
\end{eqnarray}

\noindent
Performing the summation over the fermion Matsubara frequencies 
$\widetilde{\omega}_{F,n}$ the self-consistent equation takes the form

\begin{eqnarray}
1 &=& \frac{\alpha}{4 \pi N} \int_0^\Lambda \frac{d^2 \vec{P}}{(2 \pi)^2}
\notag \\
&\Bigg(&
\frac{P \tanh \beta \sqrt{\vec{P}^2 + m(\beta)^2}}
{\left[\vec{P}^2 + \widetilde{\Pi}_3(m=0) \right] \sqrt{\vec{P}^2 
+ m(\beta)^2}} \Bigg)
\label{Mass2}
\end{eqnarray}

Eq.\eqref{Mass2} can be solved numerically with a cutoff $\Lambda$ fixed at
$\infty$ in an anlytical calculation.
By inspection of equation \eqref{Mass2} and the corresponding result obtained
by Dorey and Mavromatos \cite{DoreyMavromatos} and Lee \cite{Lee-98} one sees
that the imaginary chemical potential used which fixes rigorously one spin 
per lattice site of the original Hamiltonian \eqref{eq1}
doubles the transition temperature. This result is coherent with the results
obtained elsewhere \cite{PRB} where spinons are massless.

Since the mass can be identified with a superconducting gap one can evaluate
the parameter $r = \frac{2 m(0)}{k_B T_c}$ where $m(0)$ is the mass at zero
temperature and $T_c$ the transition temperature for which the mass becomes
zero. Dorey and Mavromatos \cite{DoreyMavromatos} obtained $r \simeq 10$ 
and Lee \cite{Lee-98} computed the mass by taking into account the frequency
dependence and obtained $r \simeq 6$. We have shown above that the imaginary
chemical potential doubles the transition temperature so that the
parameter $r$ is $ \simeq 4.8$ for $\alpha/\Lambda = \infty$ to compare
with the result of Dorey and Mavromatos and $r \simeq 3$ to compare with
Lee's result. Recall that the BCS parameter $r$ is roughly equal
to $3.5$ and the $Y Ba CuO$ parameter $r \simeq 8$ as given by the experiment
\cite{Y-Ba-Cu-O}.

\section{Conclusion}

We mapped a Heisenberg $2d$ Hamiltonian describing an antiferromagnetic quantum
spin system into a  $QED_{(2+1)}$ Lagrangian coupling a Dirac spinon field 
with a $U(1)$ gauge field. In this framework we showed that the implementation
of the constraint which fixes rigorously the site occupation in a quantum spin 
system described by a $2d$ Heisenberg model leads to a substantial 
quantitative modification of the transition temperature at which the 
dynamically generated mass vanishes in the $QED_{(2+1)}$ description. 
It modifies consequently the effective static potential which acts between 
two test particles of opposite charges.

The imaginary chemical potential \cite{Popov-88} reduces the screening of this 
static potential between test fermions when compared to the potential obtained 
from standard $QED_{(2+1)}$ calculations by Dorey and Mavromatos 
\cite{DoreyMavromatos} who implicitly used a Lagrange multiplier procedure 
in order to fix the number of particles per lattice site \cite{awe,Manousakis}
since $\lambda=0$ at the mean-field level.

We showed that the transition temperature to ``chiral'' symmetry 
restoration corresponding to the vanishing of the spinon mass $m(\beta)$ is 
doubled by the introduction of the Popov-Fedotov imaginary chemical 
potential. The trend is consistent with earlier results concerning the value 
of $T_c$ \cite{PRB}. It reduces sizably the parameter 
$r = \frac{2 m(0)}{k_B T_c}$ determined by Dorey and Mavromatos 
\cite{DoreyMavromatos} and Lee \cite{Lee-98}. 


Marston \cite{Marston} showed that in order to remove
``forbidden'' $U(1)$ gauge configuration of the antiferromagnet Heisenberg
model a Chern-Simons term should be naturally included in the $QED_3$ action 
and fix the total flux through a plaquette. When the magnetic flux through
a plaquette is fixed the system becomes 
$2\pi$-invariant in the gauge field $a_\mu$ and instantons appear in the
system. This is the case when the present non-compact formulation of $QED_3$
is replaced by its correct compact version \cite{Polya}. 

It is our next aim to implement a Chern-Simons term \cite{Dunne} in a system
constrained by a rigorous site ocupation.

\begin{acknowledgments}
One of us (R.D.) would like to thank M. Rausch de Traubenberg and J.-L. Jacquot
for interesting and fruitful discussions.
\end{acknowledgments}

\appendix
\section{Derivation of the Euclidean QED action in (2+1) dimensions 
\label{AppendixA}}

At low energy near the two independent points 
$\vec{k}= \left( \pm \pi,\pi \right) + \vec{k}$ of the Spin Brillouin Zone 
(see figure \ref{Fig1}) the Hamiltonian \eqref{HDijK} can be rewritten in
the form

\begin{eqnarray*}
H &=& \underset{\vec{k} \in SBZ}{\sum} \underset{\sigma}{\sum}
\left( f^\dagger_{1, \vec{k}, \sigma} \, f^\dagger_{1, \vec{k}
+\vec{\pi} , \sigma} \, f^\dagger_{2,\vec{k},\sigma} \, 
f^\dagger_{2,\vec{k}+\vec{\pi},\sigma} \right) 
\notag \\
&\Bigg\{&
-\mu \Unitmatrix + \sqrt{2} \Delta \left[ - k_x 
\left(
\begin{array}{cc}
\tau_3 & 0 \\
0 & \tau_3
\end{array}
\right) 
- k_y \Unitmatrix
\right]
\notag \\
&+&
\sqrt{2} \Delta \left[
-k_x
\left(
\begin{array}{cc}
\tau_2 & 0 \\
0 & -\tau_2
\end{array}
\right)
+ k_y . i \Unitmatrix
\right]
\Bigg\}
\left(
\begin{array}{c}
f_{1, \vec{k}, \sigma} \\
f_{1, \vec{k}+\vec{\pi}, \sigma} \\
f_{2,\vec{k},\sigma} \\
f_{2,\vec{k}+\vec{\pi},\sigma}
\end{array}
\right)
\end{eqnarray*}
with $\vec{\pi} = (\pi,\pi)$ the Brillouin vector. $\tau_1,\tau_2$ and 
$\tau_3$ are Pauli matrices

\begin{eqnarray*}
\tau_1 =
\left( 
\begin{array}{cc}
0 & 1 \\
1 & 0
\end{array}
\right)
, \,
\tau_2 =
\left(
\begin{array}{cc}
0 & -i \\
i & 0
\end{array}
\right)
, \,
\tau_3 =
\left(
\begin{array}{cc}
1 & 0 \\
0 & -1
\end{array}
\right)
\end{eqnarray*}

$f^\dagger_{1,\vec{k},\sigma}$ and $f_{1,\vec{k},\sigma}$ 
($f^\dagger_{2,\vec{k},\sigma}$ and $f_{2,\vec{k},\sigma}$) 
are fermion creation and annihilation operators near the point 
$(\pi,\pi)$ ($(-\pi,\pi)$).

Rotating the operators 

\begin{eqnarray*}
\begin{cases}
f_{\vec{k}} = \frac{1}{\sqrt{2}} 
\left(f_{a,\vec{k}} + f_{b,\vec{k}} \right) \\
f_{\vec{k}+\vec{\pi}} = \frac{1}{\sqrt{2}} 
\left(f_{a,\vec{k}} - f_{b,\vec{k}} \right)
\end{cases}
\end{eqnarray*}
leads to

\begin{eqnarray*}
H &=& \underset{\vec{k} \in SBZ}{\sum} \underset{\sigma}{\sum}
\psi^\dagger_{\vec{k} \sigma}
\Bigg[
- \mu \Unitmatrix 
\notag \\
&+& \widetilde{\Delta} k_{+}
\left(
\begin{array}{cc}
\tau_1 & 0 \\
0 & \tau_2
\end{array}
\right)
- \widetilde{\Delta} k_{-}
\left(
\begin{array}{cc}
\tau_2 & 0 \\
0 & \tau_1
\end{array}
\right)
\Bigg] \psi_{\vec{k} \sigma}
\end{eqnarray*}
where $k_{+} = k_x + k_y$ and $k_{-} = k_x - k_y$, 
$\widetilde{\Delta} = 2 \Delta \cos \frac{\pi}{4}$ and

\begin{eqnarray*}
\psi_{\vec{k} \sigma} = \left(
\begin{array}{c}
f_{1 a, \vec{k} \sigma} \\
f_{1 b, \vec{k} \sigma} \\
f_{2 a \vec{k} \sigma} \\
f_{2 b \vec{k} \sigma}
\end{array}
\right)
\end{eqnarray*}

In the Euclidean metric the action reads

\begin{eqnarray*}
S_{E} &=&
\int_0^\tau d\tau \underset{\vec{k} \in SBZ}{\sum} 
\underset{\sigma}{\sum}
\psi^\dagger_{\vec{k} \sigma}
\left(
\begin{array}{cc}
\tau_3 & 0 \\
0 & \tau_3
\end{array}
\right)
\Bigg[
\left(\partial_\tau - \mu \right)
\left(
\begin{array}{cc}
\tau_3 & 0 \\
0 & \tau_3
\end{array}
\right)
\notag \\
&+& i \widetilde{\Delta} k_{+}
\left(
\begin{array}{cc}
\tau_2 & 0 \\
0 & -\tau_1
\end{array}
\right)
+ i \widetilde{\Delta} k_{-}
\left(
\begin{array}{cc}
\tau_1 & 0 \\
0 & -\tau_2
\end{array}
\right)
\Bigg] \psi_{\vec{k} \sigma}
\end{eqnarray*}

\noindent
Through the unitary transformation

\begin{eqnarray*}
\psi_{\vec{k} \sigma} \rightarrow 
\left(
\begin{array}{cc}
1 & 0 \\
0 & e^{i \frac{\pi}{4} \tau_3}
\end{array}
\right)
. \left(
\begin{array}{cc}
1 & 0 \\
0 & -\tau_1
\end{array}
\right) \psi_{\vec{k} \sigma}
\end{eqnarray*}
and writing $k_{+}=k_2$ and $k_{-} = k_1$

\begin{eqnarray*}
S_{E} &=& \int_0^\beta d\tau \underset{\vec{k} \in SBZ}{\sum} 
\underset{\sigma}{\sum}
\bar{\psi}_{\vec{k} \sigma} \Big[ 
\notag \\
&&
\gamma^0 \left( \partial_\tau - \mu \right)
+ \widetilde{\Delta} i k_1 \gamma^1 + \widetilde{\Delta} i k_2 \gamma^2 \Big]
\psi_{\vec{k} \sigma}
\end{eqnarray*}

\noindent
where $\bar{\psi} = \psi^\dagger \gamma^0$ and the $\gamma$ matrices are 
defined as

\begin{eqnarray*}
\gamma^0 =
\left(
\begin{array}{cc}
\tau_3 & 0 \\
0 & -\tau_3
\end{array}
\right)
 ,\,
\gamma^1 =
\left(
\begin{array}{cc}
\tau_1 & 0 \\
0 & -\tau_1
\end{array}
\right)
 ,\,
\gamma^2 =
\left(
\begin{array}{cc}
\tau_2 & 0 \\
0 & -\tau_2
\end{array}
\right)
\end{eqnarray*}

\noindent
Using the inverse Fourier transform 
$\psi_{\vec{k} \sigma} = \int d^2\vec{r} \psi_{\vec{r} \sigma} 
e^{i \vec{k}.\vec{r}}$ the Euclidean action finally reads

\begin{eqnarray*}
S_{E} = \int_0^\beta d\tau \int d^2\vec{r} \underset{\sigma}{\sum}
\bar{\psi}_{\vec{r} \sigma} \left[ \gamma^0 \left( \partial_\tau - \mu \right)
+ \widetilde{\Delta} \gamma^k \partial_k \right] \psi_{\vec{r} \sigma}
\end{eqnarray*}

\noindent
With a ``light velocity'' 
$v_\mu = (1,\widetilde{\Delta},\widetilde{\Delta})$.
The covariant derivative which takes $v_\mu$ into account \cite{Volovik} reads

\begin{eqnarray*}
D_\mu = \partial_\mu + \frac{1}{8} \omega_{\alpha,a b} 
\left[\gamma^a,\gamma^b \right]
\end{eqnarray*}
where $\omega_{\alpha, a b} = e_a^\nu \left( \partial_\alpha e_{\nu b}
- \Gamma_{\alpha \mu}^\gamma e_{\gamma b} \right)$, $e_a^\mu$ are the 
\emph{vierbein} \cite{Ramond} for which the metric is defined as
$g^{\mu \nu} = \eta^{mn} e_m^\mu e_n^\nu = v^\mu \delta^{\mu \nu}$ with
$\eta^{00}=-1, \eta^{ij}=\delta^{ij}$ and $\Gamma$ is the Christoffel
symbols. Since $\widetilde{\Delta}$ is constant
we see clearly that the \emph{vierbein} are also constant,   
$\omega_{\alpha, a b} = 0$ in a dilated flat space-time with
the Euclidean metric $g_{\mu \nu} = v_\mu \delta_{\mu \nu}$.

\section{Derivation of the photon polarisation function at finite temperature
\label{AppendixPolarisation}}

The Fourier transformation of the spinon action given by 
Eq.\eqref{SpinonGaugeAction} reads

\begin{eqnarray*}
S_{E}\left[ \psi \right] &=& \underset{\sigma}{\sum}
\underset{\widetilde{\omega}_{F,1},\widetilde{\omega}_{F,2}}{\sum}
\int \frac{d^2\vec{k}_1}{(2 \pi)^2} \int \frac{d^2\vec{k}_2}{(2 \pi)^2}
\notag \\
\bar{\psi}_\sigma \left( k_1 \right) &\Bigg[& \frac{i \gamma^\mu k_\mu}
{(2\pi)^2 \beta} \delta \left(k_1 - k_2 \right) - \frac{i g \gamma^\mu 
a_\mu(k_1-k_2)}{(2 \pi)^2 \beta)^2} \Bigg] \psi_\sigma \left( k_2 \right)
\end{eqnarray*}
with $k = (\widetilde{\omega}_{F} \equiv \frac{2 \pi}{\beta}(n+1/4),\vec{k})$. 
Integrating over the fermion 
field $\psi$ and keeping the second order in the gauge field leads to the 
effective gauge action 

\begin{eqnarray*}
S_{eff}\left[a\right] 
= \frac{1}{2} Tr \left[G_{F}. i g \gamma^\mu a_\mu \right]^2
\end{eqnarray*}
with $Tr = \underset{\omega^{'}_{F}}{\sum} \int \frac{d^2\vec{k^{'}}}{(2\pi)^2}
. \underset{\omega^{''}_{F}}{\sum} \int \frac{d^2\vec{k^{''}}}{(2\pi)^2} 
tr$. The trace $tr$ extends over the $\gamma$ matrix space, and
$G_{F}^{-1}(k_1-k_2) = i \frac{\gamma^\mu k_\mu}{(2\pi)^2\beta} 
\delta(k_1-k_2)$. The pure gauge action comes as

\begin{eqnarray*}
S_{eff}\left[ a \right] = - g^2 \frac{1}{2 \beta} \underset{\sigma}{\sum} 
\underset{\omega_{F,1}}{\sum} \int \frac{d^2\vec{k}_1}{(2\pi)^2}.
\frac{1}{\beta} \underset{\omega_{F}^{''}}{\sum} \int \frac{d^2\vec{k^{''}}}
{(2\pi)^2}
\notag \\
tr \Bigg[
\frac{\gamma^\rho k_{1,\rho}}{k_1^2}.\gamma^\mu a_\mu(k_1-k^{''}).
\frac{\gamma^{\eta} k^{''}_\eta}{{k^{''}}^2}.
\gamma^\nu a_\nu \left(-(k_1 - k^{''}) \right)
\Bigg]
\end{eqnarray*}

\noindent
With the change of variables $k_1-k^{''}=q$ and $k_1 = k$

\begin{eqnarray*}
S_{eff} =
- \frac{g^2}{2 \beta}  
\underset{\omega_{B}}{\sum} \int \frac{d^2\vec{q}}{(2\pi)^2}
a_\mu(-q) \Pi^{\mu \nu}(q) a_\nu(q). 
\end{eqnarray*}
where $q = (\omega_{B}=\frac{2\pi}{\beta} m, \vec{q})$ and
the polarisation function is given by

\begin{eqnarray*}
\Pi^{\mu \nu}(q) = \frac{g^2}{\beta} \underset{\sigma}{\sum}  
\underset{\omega_{F}}{\sum} \int \frac{d^2\vec{k}}{(2\pi)^2}
tr \left[
\frac{\gamma^\rho k_\rho}{k^2} . \gamma^\mu. \gamma^\eta
\frac{\left(k_\eta + q_\eta \right)}{\left(k + q \right)}. \gamma^\nu
\right]
\end{eqnarray*}

Then using the Feynmann identity $\frac{1}{a b} = \int_0^1 dx 
\frac{1}{\left(a x + (1-x)b \right)^2}$ $\Pi^{\mu \nu}$ can be rewritten as

\begin{eqnarray*}
\Pi^{\mu \nu}(q) &=& \frac{g^2 }{\beta} \underset{\sigma}{\sum}  
\underset{\omega_F}{\sum} \int \frac{d^2\vec{k}}{(2\pi)^2}
tr \left[\gamma^\rho \gamma^\mu \gamma^\eta \gamma^\nu \right].
\notag \\
&& \int_0^1 dx \frac{k_\rho (k_\eta + q_\eta )}
{\left[(k+q)^2 x + (1-x) k^2 \right]^2}
\end{eqnarray*}

By means of a change of variables
$k \rightarrow k^{'} - x q$ and using the identity 
$tr \left[\gamma^\rho \gamma^\mu \gamma^\eta \gamma^\nu \right]=
4.\left[\delta_{\rho \mu}.\delta_{\eta \nu}-
\delta_{\rho \eta}.\delta_{\mu \nu} + \delta_{\rho \nu}.\delta_{\mu \eta} 
\right]$ one obtains

\begin{eqnarray*}
\Pi^{\mu \nu}(q) &=& 4 \alpha \int_0^1 dx \frac{1}{\beta}
\underset{\omega_F^{'}}{\sum} \int \frac{d^2\vec{k^{'}}}{(2\pi)^2}
\Bigg\{
\Big[
2 k^{'}_\mu k^{'}_\nu 
\notag \\
&+& (1-2x)(k^{'}_\mu q_\nu + q_\mu k^{'}_\nu)
-x(1-x) 2 q_\mu q_\nu 
\notag \\
&-& \delta_{\mu \nu} \underset{\eta}{\sum} \left( 
{k^{'}_\eta}^2 + (1-2x) k^{'}_\eta q_\eta - x(1-x) {q_\eta}^2
\right)
\notag \\
&& \Big]/\left[{k^{'}}^2 + x(1-x)q^2 \right]^2
\Bigg\}
\end{eqnarray*}
where $\alpha = g^2 \overset{N=2}{\underset{\sigma = 1}{\sum} 1 }$.
Following Dorey and Mavromatos \cite{DoreyMavromatos}, Lee \cite{Lee-98},
Aitchison \emph{et al.} \cite{Aitchison-92} and Gradshteyn \cite{Gradshteyn}
we define

\begin{eqnarray*}
S_1 &=& \overset{\infty}{\underset{n = -\infty}{\sum}}
\frac{1}{\left[{k^{'}}^2 + x(1-x)q^2 \right]}
\notag \\
&=&
\frac{\beta^2}{4 \pi Y} \left[ \frac{\sinh (2\pi Y)}{\cosh(2\pi Y)
-\cos(2\pi X)} \right]
 \\
S_2 &=& \overset{\infty}{\underset{n = -\infty}{\sum}}
\frac{1}{\left[{k^{'}}^2 + x(1-x)q^2 \right]^2}
= -\frac{\beta^2}{8 \pi^2}.\frac{1}{Y} \frac{\partial S_1}{\partial Y}
 \\
S^{*} &=& \overset{\infty}{\underset{n = -\infty}{\sum}}
\frac{\omega^{'}_F}{\left[{k^{'}}^2 + x(1-x)q^2 \right]^2}
= - \frac{\beta}{4\pi} \frac{\partial S_1}{\partial X}
\end{eqnarray*}
with $X= x.m + 1/4$ and $Y = \frac{\beta}{2\pi} 
\sqrt{\vec{k^{'}}^2 + x(1-x)q^2}$.
The polarisation can be expressed in terms of these sums and reads

\begin{eqnarray*}
\Pi^{00} &=& \frac{\alpha}{\beta} \int_0^1 dx 
\int \frac{d^2\vec{k^{'}}}{(2\pi)^2}
\Big[ S_1 - 2\left[\vec{k^{'}}^2 + x(1-x)q_0^2 \right]S_2 
\notag \\
&+& (1-2x)q_0 S^{*} \Big]
\end{eqnarray*}
for the temporal component and

\begin{eqnarray*}
\Pi^{ij} &=& \frac{\alpha}{\beta} \int_0^1 dx 
\int \frac{d^2\vec{k^{'}}}{(2\pi)^2}
\Big[ 2x(1-x)(q^2 \delta_{ij} - q_i q_j)S_2 
\notag \\
&-& (1-2x) q_0 \delta_{ij}. S^{*} \Big]
\end{eqnarray*}
for the spatial components.

Integrating over the fermion momentum $\vec{k^{'}}$ one gets

\begin{eqnarray*}
\Pi^{00} &=& \widetilde{\Pi}_3 - \frac{q_0^2}{q^2} \widetilde{\Pi}_1 - 
\widetilde{\Pi}_2
 \\
\Pi^{ij} &=& \widetilde{\Pi}_1 \left( \delta_{ij} - \frac{q_i q_j}{q^2} \right)
+ \widetilde{\Pi}_2 \delta_{ij}
\end{eqnarray*}
where

\begin{eqnarray*}
\widetilde{\Pi}_1 &=& \frac{\alpha q}{\pi} \int_0^1 dx \sqrt{x(1-x)}
\frac{\sinh \beta q \sqrt{x(1-x)} }{D(X,Y)}
 \\
\widetilde{\Pi}_2 &=& \frac{\alpha m}{\beta}
\int_0^1 dx (1-2x) \frac{\cos 2 \pi x m}{D(X,Y)}
 \\
\widetilde{\Pi}_3 &=& \frac{\alpha}{\pi \beta} \int_0^1 dx \log 2 D(X,Y)
\end{eqnarray*}
and $D(X,Y)= \cosh \left( \beta q \sqrt{x(1-x)} \right) + \sin (2\pi x m) $.

\end{document}